\documentclass[a4paper,10pt,twocolumn]{article}

\usepackage[utf8]{inputenc}
\usepackage[square,comma,numbers]{natbib}
\usepackage[french]{babel}
\usepackage{graphicx,times}
\usepackage[T1]{fontenc}
\usepackage{amsmath}
\usepackage[font=footnotesize]{caption}
\usepackage{fancyhdr}
\usepackage[explicit]{titlesec}
\usepackage{hyperref}
\usepackage{tabularx}

\newcolumntype{L}[1]{>{\raggedright\arraybackslash}p{#1}}
\newcolumntype{C}[1]{>{\centering\arraybackslash}p{#1}}
\newcolumntype{R}[1]{>{\raggedleft\arraybackslash}p{#1}}

\hypersetup{colorlinks=true, urlcolor=blue, urlbordercolor={0 0 1}, citecolor=black, citebordercolor={1 1 1}}

\addto\captionsfrench{}
\addto\captionsfrench{}

\titleformat{\section}
  {\normalfont}{\thesection.}{0.5em}{\MakeUppercase{#1}}
\titleformat{\subsection}
  {\normalfont\itshape}{\thesubsection.}{1.5em}{#1}
\titleformat{\subsubsection}
  {\normalfont\itshape}{\thesubsubsection.}{1.5em}{#1}
	
\titlespacing\section{0pt}{1em}{0.5em}
\titlespacing\subsection{0pt}{1em}{0.5em}
\titlespacing\subsubsection{0pt}{1em}{0.5em}

\title{
\fontsize{24pt}{24pt}\selectfont
Commande rapprochée d'un IGBT pour l'atténuation des perturbations électromagnétiques
}

\author{
\fontsize{11pt}{11pt}\selectfont
Daniel Sting Martinez-Padron\textsuperscript{1}, Nicolas Patin\textsuperscript{1} et Eric Monmasson\textsuperscript{2} \\
\fontsize{10pt}{10pt}\selectfont
\textsuperscript{1}Université de technologie de Compiègne, Roberval (Mechanics, energy and electricity), Centre de recherche\\
\fontsize{10pt}{10pt}\selectfont
Royallieu  –  CS 60319  – 60203 Compiègne Cedex – France.\\
\fontsize{10pt}{10pt}\selectfont
\textsuperscript{2}CY Cergy Paris Université, SATIE laboratory, Cergy-Pontoise, France. 
}

\date{}

\begin{document}

\maketitle
\thispagestyle{fancy}

\fontsize{9pt}{9pt}\selectfont
\textbf{RESUME --Les transistors de puissance comme les IGBT et les MOSFET sont une source d'interférences électromagnétiques (EMI) pendant les commutations en raison des variations rapides de tension/courant. L'augmentation de la durée des commutations peut réduire la production des EMI mais augmente les pertes. Plusieurs techniques de commande rapprochée pour réduire ces EMI ont été proposées dans la littérature. Dans ce travail, une commande basée sur le contrôle d'un profil de courant de grille, rendu possible par de nouveaux drivers disponibles sur le marché, est proposé. Pour cela, on s'appuie sur la caractéristique du transistor de tension grille-source en fonction de la charge injectée dans la grille. Afin de démontrer la performance de cette méthode, elle est évaluée par simulation SPICE à l'aide d'un facteur de mérite permettant de la comparer quantitativement à une technique de référence dénommée CATS (Commande autour de la Tension de Seuil).}\\

\textbf{\textit{Mots-clés -- IGBT, commande rapprochée, EMI, Pertes.}}

\fontsize{10pt}{10pt}\selectfont

\section{Introduction}
Les transistors de puissance comme les IGBT et les MOSFET sont largement utilisés dans de nombreuses applications telles que les variateurs de vitesse industriels, les groupes moto-propulseurs des véhicules électriques/hybrides ou les applications grand public. Cependant, un inconvénient de ces dispositifs est qu'ils génèrent des perturbations électromagnétiques (EMI) pendant les commutations car les niveaux élevés de $dv/dt$ et $di/dt$ produisent des EMI conduites et rayonnées \cite{bib_1}. Néanmoins, ces EMI peuvent être réduites en augmentant la durée des commutations, mais cela tend à augmenter les pertes dans les semiconducteurs. On notera que la durée et la forme des $dv/dt$ et $di/dt$ sont liées au courant injecté dans la grille. Le moyen le plus simple (et le plus classique) de ralentir la commutation des transistors est d'intercaler entre le driver (source de tension quasi idéal) et la grille (capacitive) du transistor une résistance qui augmentera la constante de temps d'évolution de la charge stockée dans la grille. Toutefois, le choix d'une résistance de grille relève d'un compromis entre les pertes et les EMI mais cette technique ne garantit pas l'optimalité des EMI pour un niveau de pertes donné. Pour cela, un contrôle plus fin de la commutation est nécessaire pour agir sur sa forme indépendemment du temps requis pour la commutation.
Pour ce faire, une première solution consiste en un driver de commande basé sur une source de courant externe et la détection du plateau de Miller, présenté dans \cite{bib_2,bib_3,bib_4}. Celle-ci permet d'obtenir un rapport adéquat entre la génération d'EMI et les pertes. Dans cette approche, un circuit push-pull est utilisé pour modifier la valeur de la résistance de la grille entre l'amorçage et le blocage du transistor. La valeur de la résistance de la grille pendant l'amorçage est choisie pour réduire la génération d'EMI en augmentant la durée du $di/dt$. La chute de tension est accélérée par l'injection d'un courant provenant d'une source externe qui est activée par un circuit de détection du plateau de Miller afin de compenser les pertes. Lors du blocage, la résistance est sélectionnée pour réduire autant que possible les pertes par commutation induites par la queue de courant, classique pour un transistor IGBT.\\
Un autre circuit de commande, basé sur une boucle fermée, est rapporté dans \cite{bib_5} et \cite{bib_6}. Un correcteur proportionnel-intégral (PI) contrôle le courant de grille injecté dans l'IGBT pour asservir  le $di/dt$ et le $dv/dt$. Il faut alors noter que la principale difficulté avec cette solution est d'obtenir un temps de réponse compatible avec les formes de consignes imposées et les temps de commutations attendus pour satisfaire aux contraintes de pertes.\\
Dans \cite{bib_7} et \cite{bib_8}, un circuit de contrôle de capacité Miller externe est ajouté afin de contrôler le $dv/dt$. Là encore, l'objectif est de contrôler la forme de $di/dt$ et $dv/dt$ avec un lissage des fronts de commutation.\\
La méthode CATS (Commande Autour de la Tension de Seuil) est proposée dans \cite{bib_9} pour réduire la génération d'EMI en lissant une fois encore les formes d'ondes de tension et de courant en appliquant un niveau de tension de grille intermédiaire, qui comme son nom l'indique se trouve au voisinage du $V_{geth}$ du transistor.\\
Il est notamment prouvé dans \cite{bib_10} qu'une forme gaussienne (ou plutôt des fronts obtenus avec l'intégrale d'une gaussienne) minimise les EMI indépendamment du temps de commutation.\\
Récemment, des circuits de pilotage \cite{bib_11}-[15] basés sur une source de courant de grille avec une résolution fine ont été mis sur le marché. Dans cet article, une méthode de pilotage basée sur un profil de courant de grille permettant de contrôler la forme de $dv/dt$ est proposée. Elle consiste à appliquer un profil de courant constant par intervalles pour lequel on sera capable de fixer l'amplitude et la durée de chacun de ces intervalles. Ces paramètres sont ajustés en prenant en compte le comportement non-linéaire du transistor au début du plateau de Miller. Cette méthode est décrite dans la section 2 et les résultats des simulations obtenues sont présentés dans la section 3. Une comparaison entre cette technique et la commande CATS est effectuée à la section 4 pour ensuite aboutir à la conclusion de l'article. 

\section{ Méthode proposée}
La méthodologie proposée est basée sur la charge de la grille du transistor qui détermine la forme et la durée des $dv/dt$ et $di/dt$ entre collecteur et émetteur. Le comportement du transistor tout au long de la charge de la grille du est décrite dans le paragraphe suivant.

\subsection{Charge de la grille du transistor}
\begin{figure}[!ht]
	\begin{center}
		\includegraphics[width=0.7\columnwidth]{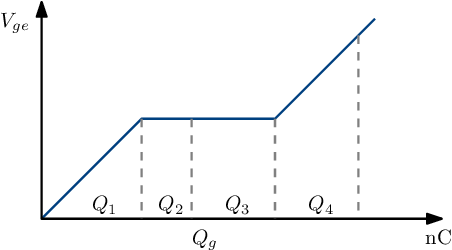}
	\end{center}
	
	\caption{Courbe de tension $V_{ge}$ en fonction de la charge injectée dans la grille.}
	\label{fig_1}
\end{figure}

\begin{figure}[!ht]
	\begin{center}
		\includegraphics[width=0.7\columnwidth]{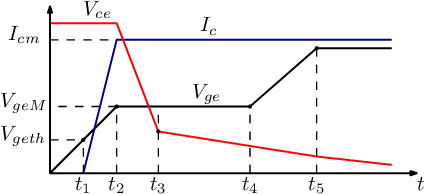}
	\end{center}
	\caption{Courbe dynamique de tension et courant dans un IGBT durant la charge de grille.}
	\label{fig_2}
\end{figure}

La courbe typique de tension grille-émetteur en fonction de la charge injectée dans la grille d'un transistor IGBT (ou MOSFET) est montrée dans la Figure \ref{fig_1}. Elle peut être divisée en plusieurs parties liées aux différents régimes de fonctionnement du transistor.
Cette caractéristique, fournie par les constructeurs, peut être obtenue en pratique en injectant un courant constant dans la grille. La croissance de la valeur de la charge $Q_g$ sera donc linéaire en fonction du temps et par ailleurs, le courant constante permet également de s'affranchir des effets des inductances parasites de la connexion entre le transistor et la source \cite{bib_16}. La courbe de la Figure \ref{fig_2} montre ensuite la dynamique du courant et de la tension entre collecteur et émetteur durant cette charge de la grille à courant constant. Cette courbe peut être divisée en 4 étapes:
\begin{itemize} 
\item \textbf{Étape 1}: $0<t<t_1$. La tension de grille $V_{ge}$ est inférieure à la tension de seuil $V_{geth}$. Le courant de collecteur $I_c$ est fortement limité (seulement quelques µA). On considère alors que le transistor est bloqué.
\item \textbf{Étape 2}: $t_1<t<t_2$. Une fois que $V_{geth}$ est atteint, le transistor est mis en conduction et le courant de collecteur circule jusqu'à atteindre sa valeur maximale $I_{cm}$ (fixée par le courant dans la charge). Durant toute cette phase, le transistor fonctionne en régime linéaire et dissipe donc beaucoup de puissance.
\item \textbf{Étape 3}: $t_2<t<t_4$. Pendant le plateau de Miller, $V_{ge}=V_{geM}$ à cause de la capacité grille-collecteur (capacité de Miller). $V_{ce}$ commence à chuter avec différentes pentes déterminées pour differentes valeurs de charge de la capacité de Miller \cite{bib_16} (d'où un découpage en deux phases avec un instant $t_3$ intermédiaire). En fait, le courant de grille ne sert plus à charger la capacité grille-émetteur ($C_{ge}$) mais à décharger la capacité grille-collecteur (d'où la chute du potentiel de collecteur par rapport à la grille mais aussi par rapport à l'émetteur).
\item \textbf{Étape 4}: $t_4<t<t_5$. Dès que l'effet de Miller se termine, la tension de grille recommence à augmenter vers sa valeur finale. Le transistor tend alors vers son régime de saturation avec des pertes qui se stabilisent au niveau des seules pertes par conduction.
\end{itemize}
 
\subsection{Profil de courant de grille proposé}
\begin{figure}[!ht]
	\begin{center}
		\includegraphics[width=0.7\columnwidth]{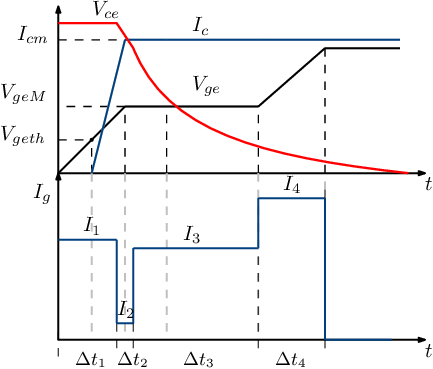}
	\end{center}
	
	\caption{Profil de courant de grille proposé (CSP).}
	\label{fig_3}
\end{figure}

Comme indiqué précédemment, le comportement transitoire de commutation, qui détermine la génération des EMI et les pertes, est fortement lié à la charge de la grille, elle-même contrôlée par le courant injecté par le driver. Pour cette raison, dans cet article, il est proposé de piloter la grille par un profil d'échelons de courant (\emph{Current Step Profile} - CSP) de telle sorte qu'il soit possible de réguler chaque étage de charge et de pouvoir contrôler le transitoire de tension. 
Ainsi pendant l'amorçage du transistor, le CSP doit permettre de ralentir la chute de tension $v_{ce}$ au début de la plateau de Miller. Le ralentissement de cette transition permet un contrôle en $dv/dt$ plus conforme à l'objectif qui est l'intégrale d'une gaussienne. Le CSP est présenté dans la Figure \ref{fig_3}: il est formé (pour des raisons de recherche de modération de la complexité) de 4 échelons de courant d'amplitude $I_j$ avec $1\leq j \leq 4$ et durée $\Delta t_j$ et chacun permet d'effectuer différentes étapes de la charge de grille. 
L'amplitude de chaque échelon est calculée en considérant les étapes de la courbe dynamique présentée dans la Figure \ref{fig_2} comme suit :
\begin{itemize}
\item  \textbf{Échelon 1.} Ce première échelon sert à réaliser les deux premières étapes de la charge du transistor, à savoir,  
premièrement le courant pour atteindre la tension de seuil du l'IGBT et deuxièmement la mise en conduction du transistor. Pendant l'étape 1, le courant de grille est déterminé par la charge $C_{ge}$ et par : 
\begin{equation}
{I_{g1}} = {C_{ge}}\frac{{d{V_{ge}}}}{{dt}} \approx {C_{ge}}\frac{{{V_{gth}}}}{{\Delta {t_{s1}}}},
\label{eq1}
\end{equation}
où $I_{g1}$ est le courant de grille pour l'étape 1, $\Delta t_{s1}$ la durée de temps pour que $V_{ge}=V_{geth}$. Durant l'étape 2, $I_c$ augmente jusqu'à $I_{cm}$ pendant $\Delta t_{s2}$ et $C_{gc}$ est chargé par le courant :
\begin{equation}
I = {C_{gc}}\frac{{d{V_{gc}}}}{{dt}}.
\label{eq2}
\end{equation}
En considérant que $V_g=V_{ge}-V_{ce}$, avec $V_{ce}$ constant et que la transconductance de l'IGBT est $g_m= \Delta I_c/\Delta V_{ge}$, le courant de grille est calculé par :
\begin{equation}
{I} = \frac{{{C_{ies}}}}{{{g_m}}}\frac{{d{I_c}}}{{dt}} \approx \frac{{{C_{ies}}}}{{{g_m}}}\frac{{\Delta {I_c}}}{{\Delta {t_{s2}}}},
\label{eq3}
\end{equation}
où $I_{g2}$ est la courant de grille pour l'étape 2. Pour la méthode proposée, $\Delta I_c$ est considéré comme une proportion du courant maximal tel que $\Delta I_c=\alpha I_{cm}$ avec $0<\alpha \leq 1$. En prenant $I_1=I_{g1}=I_{g2}$, la durée de l'échelon 1, $\Delta t_1$ est calculée à partir des équations (\ref{eq1}) et (\ref{eq3}) et $\Delta t_1=\Delta t_{s1}+\Delta t_{s2}$.
\item  \textbf{Échelon 2.} Cet échelon contrôle la fin du transitoire de courant et la transition entre $di/dt$ et $dv/dt$ (début du plateau de Miller). Afin d'avoir un contrôle adéquat du $dv/dt$ et en raison du comportement non-linéaire de la capacité d'entrée et de l'effet de la récupération de la diode, dans cette méthodologie, il est proposé de ralentir cette transition avec un échelon de faible amplitude $I_2$ dont la durée $\Delta t_2$ est calculée avec l'équation (\ref{eq3}).
\item  \textbf{Échelon 3.} Pendant cette étape, le $dv/dt$ est contrôlé. A partir de l'échelon 3. Le courant de grille peut être estimé par :
\begin{equation}
{I_3} = {C_{gc}}\frac{{d{V_{ce}}}}{{dt}} \approx {C_{gc}}\frac{{\Delta {V_{ce}}}}{{\Delta {t_3}}}.
\end{equation}
En fixant la durée  $\Delta t_3$, le $dv/dt$ est contrôlé par l'amplitude de l'échelon 3. 
\item  \textbf{Échelon 4.} Afin de finir la charge de grille, l'amplitude de l'échelon $I_4$  est calculée en utilisant (\ref{eq1}) et en considérant la capacité équivalente pendant ce temps.
\end{itemize}

\section{Évaluation}
\begin{figure}[!ht]
	\begin{center}
		\includegraphics[width=0.7\columnwidth]{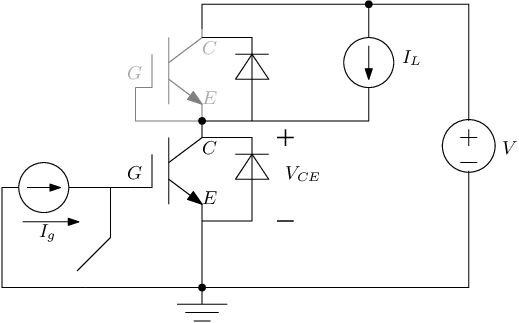}
	\end{center}
	
	\caption{Circuit de test pour l'évaluation du CSP.}
	\label{fig_4}
\end{figure}

Le CSP est évalué par simulation en utilisant le logiciel Ltspice et le modèle SPICE de l'IGBT IKW40N65ET7. Le circuit de test est présenté à la Figure \ref{fig_4}. La source de courant générant le CSP est paramétré avec quatre amplitudes ($I_1$, $I_2$, $I_3$ et $I_4$) successives et leurs durées associées ($\Delta t_1$, $\Delta t_2$, $\Delta t_3$ et $\Delta t_4$). Tous les paramètres sont fixés à l'exception de $I_3$ qui est celui qui va permettre de contrôler la durée de chute de la tension $V_{ce}$. On fait alors varier sa valeur afin d'évaluer son impact sur la génération des EMI et les pertes. Les simulations ont été réalisées avec un pas minimal de 100 ps et les paramètres électriques du circuit sont résumés dans le Tableau \ref{tab_1}. 
\begin{table}[!ht]
	\caption{Paramètres d'essai.}
	
	\begin{center}
		\begin{tabular}{|>{\footnotesize}L{3.0cm}|>{\footnotesize}L{1.5cm}|>{\footnotesize}L{2.4cm}|>{\footnotesize}L{2.1cm}|}
			\hline
			\textbf{Paramètre}&\textbf{Valeur}\\
			\hline
	Capacité d'entrée ($C_{ies}$)  & 2.475 nF\\ \hline
	Capacité de transfert inverse ($C_{res}$)  & 25 pF\\ \hline
	Transconductance ($g_{m}$)  & 21 S\\ \hline
	Tension de seuil ($V_{geth}$) & 5.8 V\\ \hline
	Tension d'entrée ($V$) & 130 V \\ \hline
    Courant de collecteur ($I_c$) & 10 A\\   
	 \hline
	 Période de commutation ($T_s$) & $50 \:\mathrm{\mu s}$\\ \hline
		\end{tabular}
	\end{center}
	
	\label{tab_1}
\end{table}
La sélection de chaque échelon du CSP est décrite comme suit.
\begin{itemize}
\item \textbf{Échelon 1}. Tout d'abord, en considérant la valeur de $V_{geth}$ et $C_{ies}$ montrée dans le Tableau \ref{tab_1} et en proposant $t_{s1}=100\:\mathrm{ns}$, le courant de grille pour atteindre $V_{geth}$ est calculé par:
\begin{equation}
I_{1}= C_{ies}\frac{V_{geth}}{\Delta t_{s1}}=143,5\:\mathrm{mA}.
\end{equation}
Pour la deuxième partie, afin de ralentir la pente du courant avant d'atteindre le plateau de Miller, il est proposé d'atteindre 80\% de $I_{cm}$, à savoir $\alpha=0,8$. Car $I_{1}=143,5\:\mathrm{mA}$, la durée est donnée par
\begin{equation}\label{eq:IG2} 
\Delta t_{s2}=\frac{C_{ies}}{g_m}\frac{\alpha I_{cm}}{I_{1}}= 25,3\:\mathrm{ns}.
\end{equation}
Par conséquent, l'amplitude d'échelon 1 est de $I_{1}=143.5\:\mathrm{mA}$ avec une durée de $\Delta t_1=125.3\:\mathrm{ns}$.
\item \textbf{Échelon 2}. Afin de terminer la pente du courant et de commencer le plateau de Miller, un échelon de faible amplitude $I_2$ est proposé. Dans ce cas $I_2=14.3\:\mathrm{mA}$ est sélectionné, ce qui correspond à 10\% de la valeur de $I_1$. La valeur de $C_{ies}$ estimée est $250\:\mathrm{pF}$, alors la durée est calculée par
\begin{equation}
\Delta t_{2}=\frac{C_{ies}}{g_m}\frac{(1-\alpha) I_{cm}}{I_{2}}=63,3\:\mathrm{ns}.
\end{equation}
L'amplitude d'échelon 2 est  $I_2=1.5\:\mathrm{mA}$ avec un durée de $\Delta t_{2}=63,3\:\mathrm{ns}$.
\item  \textbf{Échelon 3}. Pour cet échelon, cinq valeurs différentes sont sélectionnées : $I_{3}=10\:\mathrm{mA}$, $I_{3}=20\:\mathrm{mA}$, $I_{3}=30\:\mathrm{mA}$, $I_{3}=40\:\mathrm{mA}$ et $I_{3}=50\:\mathrm{mA}$. La durée est fixée à $\Delta t_{3}=400\:\mathrm{ns}$ pour tous les cas.
\item \textbf{Échelon 4}. 
Afin de terminer la charge de grille, la durée proposée est $\Delta t_{4}=500\:\mathrm{ns}$, la capacité estimée à partir de la courbe de charge de grille de la documentation du composant est $C=20\:\mathrm{nF}$ et l'amplitude est calculée par
\begin{equation}
I_{4}= C\frac{\Delta V}{\Delta t_{4}}=296\:\mathrm{mA}.
\end{equation}
L'amplitude du pas 4 est de $I_4=296\:\mathrm{mA}$ avec une durée de $\Delta t_{4}=500\:\mathrm{ns}$.
\end{itemize}
\begin{figure}[!ht]
	\begin{center}
		\includegraphics[width=0.9\columnwidth]{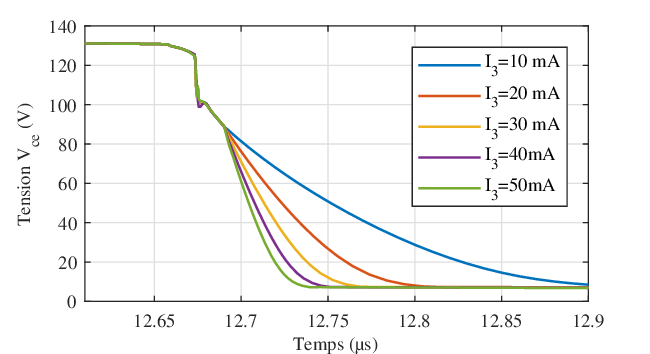}
	\end{center}
		\caption{Résultats des simulations pour différentes valeurs de $I_3$ pour le front descendant.}
	\label{fig_5}
\end{figure}

\begin{figure}[!ht]
	\begin{center}
		\includegraphics[width=0.9\columnwidth]{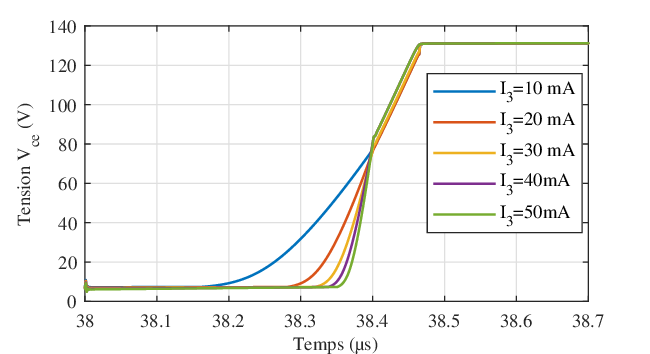}
	\end{center}
		\caption{Résultats des simulations pour différentes valeurs de $I_3$ pour front montant.}
	\label{fig_6}
\end{figure}

\begin{figure}[!ht]
	\begin{center}
		\includegraphics[width=0.9\columnwidth]{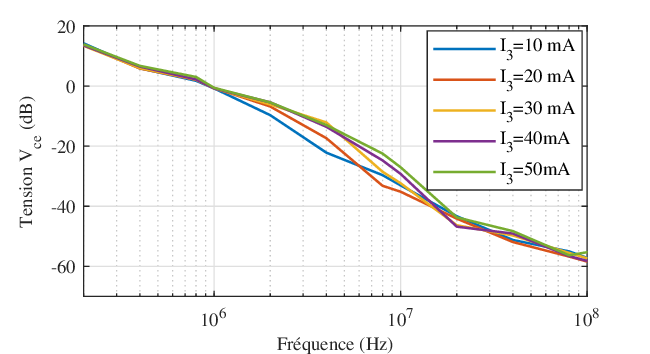}
	\end{center}
		\caption{Spectres de fréquence à différents niveaux de $I_3$.}
	\label{fig_7}
\end{figure}
Les résultats obtenus dans les Figures \ref{fig_5} et \ref{fig_6} montrent la forme d'onde $V_{ce}$ pendant le front descendant et montant à différentes amplitudes de $I_3$. Les  spectres pour chaque à la Figure \ref{fig_7}  montrent que la génération des EMI peut être réduite en diminuant l'amplitude de $I_3$. Les pertes par commutations générées sont calculées comme indiqué dans la documentation du transistor \cite{bib_17} : 
\begin{itemize}
    \item Pour la mise en conduction, les pertes par commutations sont mesurées pendant l'intervalle de temps entre 10 \% de $V_{ge\max} = 15 \:\mathrm{V}$ et 2\% de $V_{ce\max} = 130\:\mathrm{V}$.
    \item Pour le blocage, les pertes par commutations sont mesurées pendant l'intervalle de temps entre 90 \% de $V_{ge\max}$ et 2\% de $I_{cm}=10\:\mathrm{A}$.
\end{itemize}
   Les pertes globales, regroupées dans le Tableau \ref{tab_2}, tendent à diminuer lorsque l'amplitude de $I_3$ augmente car cette augmentation accélère le transitoire de tension du transistor.\\
   Du point de vue des perturbations électromagnétiques, un facteur de mérite (FOM) introduit dans \cite{bib_19} a été utilisé. Celui-ci est basé sur l'inégalité de Heisenberg-Gabor (HG) et le co-étalement temps-fréquence des signatures des fronts montants et descendants (le signal réel pouvant être obtenu par un produit de convolution entre le signal MLI idéal - à commutations instantanées - et un signal "signature" $\lambda(t)$ de la forme de la commutation réelle). C'est ce signal $\lambda(t)$ (en fait deux signaux, l'un décrivant les fronts montants et l'autre les fronts descendants) qui est analysé pour évaluer son étalement temporel $\sigma_t$ puis, après transformée de Fourier, on évalue également son étalement fréquentiel $\sigma_\omega$. L'inégalité HG montre que quelle que soit la fonction $\lambda(t)$, on a toujours $\sigma_t.\sigma_\omega \le \frac{1}{2}$ et que la borne est atteinte pour un $\lambda(t)$ gaussien. Le FOM présenté étant la somme des co-étalements des $\lambda(t)$ pour les fronts montant et descendant, le cas idéal théorique serait un FOM = 1. On peut donc voir que cette borne est loin d'être atteindre ici mais qu'il existe un optimum de réglage $I_3=30 mA$ permettant de le minimiser. Par contre, cela se fait au prix d'une légère dégradation des pertes par rapport à $I_3=50 mA$ avec une augmentation de 6.8\% des pertes (mais dans le cas contraire, le choix optimal pour les pertes occasionne une dégradation de 23.5\% du FOM et donc des perturbations électromagnétiques).
\begin{table}[!ht]
	\caption{Résultats de l'évaluation.}
	
	\begin{center}
		\begin{tabular}{|>{\footnotesize}L{1.0cm}|>{\footnotesize}L{1.8cm}|>{\footnotesize}L{1.5cm}|>{\footnotesize}L{1.1cm}|}
			\hline
			\textbf{$I_3$}&\textbf{Pertes par commutation}&\textbf{FOM}\\
			\hline
	10 mA & 7,95 W  & 181,45\\ \hline
	20 mA & 6,58 W  & 179,73\\ \hline
	30 mA & 6,10 W  & 170,12\\ \hline
	40 mA & 5,95 W  & 184,31\\ \hline
	50 mA & 5,71 W  & 210,76\\ \hline
	 
		\end{tabular}
	\end{center}
	
	\label{tab_2}
\end{table}

\section{Comparaison}
La performance du CSP est évaluée par une comparaison avec une méthode efficace rapportée dans la littérature, à savoir la méthodologie CATS.
\subsection{Méthodologie CATS}
\begin{figure}[!ht]
	\begin{center}
		\includegraphics[width=0.7\columnwidth]{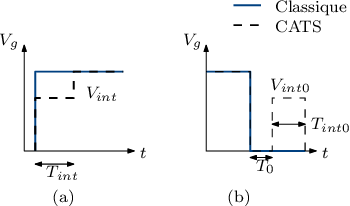}
	\end{center}
		\caption{Principe de fonctionnement de CATS : (a) amorçage et (b) blocage.}
	\label{fig_8}
\end{figure}

La méthodologie CATS est proposée dans \cite{bib_9} et consiste à réduire les $di/dt$ et $dv/dt$ en débutant l'amorçage du transistor à l'aide d'un générateur de tension $V_{g}$ que l'on fixe à une valeur intermédiaire $V_{int}$ proche (mais légèrement supérieure) à la tension $V_{geth}$ pendant une durée $T_{int}$ comme le montre la Figure \ref{fig_8}(a) contrairement au pilotage classique. 

Lors du blocage, une tension nulle ou négative est forcée pour évacuer rapidement la charge stockée, qui maintient le transistor en conduction pendant un intervalle de temps $T_o$. Puis un deuxième niveau $V_{int0}<V_{int}$ est injecté pendant $T_{int0}$ comme montré dans la Figure \ref{fig_8}(b).
\subsection{Comparaison et résultats}
\begin{figure}[!ht]
	\begin{center}
		\includegraphics[width=0.9\columnwidth]{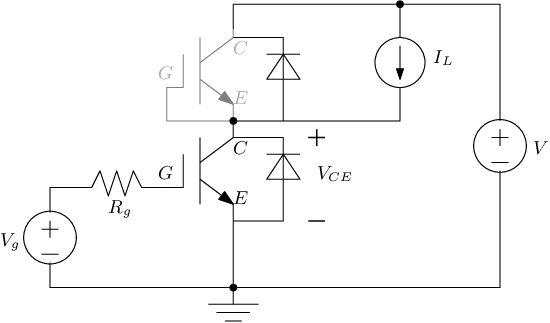}
	\end{center}
		\caption{Principe de fonctionnement de CATS : (a) amorçage et (b) blocage.}
	\label{fig_9}
\end{figure}

La comparaison consiste à déterminer les performances de génération d'EMI des deux méthodes (CSP et CATS) pour un même niveau de pertes par commutations. Elle est effectuée en utilisant le circuit de test montré dans la Figure \ref{fig_9} avec l'IGBT IKW40N65ET7 et le logiciel Ltspice en prenant toujours un pas de simulation maximum de 100 ps. Le niveau de pertes par commutations choisi est $6,58 W$. Les paramètres du système sont: $V$=130 V, $T_s$= $50 \:\mathrm{\mu s}$ et $I_c$=10 A.
Pour la méthodologie CATS, les paramètres $T_{int}$, $V_{int0}$, $T_0$ et $V_{int}$ sont choisi comme est indiqué dans le Tableau \ref{tab_3}.
\begin{table}[!ht]
	\caption{Paramètres pour CATS.}
	
	\begin{center}
		\begin{tabular}{|>{\footnotesize}L{1.5cm}|>{\footnotesize}L{1.5cm}|>{\footnotesize}L{1.8cm}|>{\footnotesize}L{1.1cm}|}
			\hline
			\textbf{Paramètre}&\textbf{Valeur}\\
			\hline
	$T_{int}$& $400$ ns \\ \hline
	$T_0$& $200$ ns \\ \hline
	$T_{int0}$& $400$ ns \\ \hline	
	$V_{int}$& $7,5$ V \\ \hline
	$V_{int0}$& $3,75$ V \\ \hline
		\end{tabular}
	\end{center}
	
	\label{tab_3}
\end{table}
Pour le CSP, la source $V_g$ dans \ref{fig_9} est remplacée par une source de courant $I_g(t)$ et la durée et l'amplitude de l'échelon $I_3$ deviennent les degrés de liberté pour ajuster les pertes par commutations. L'amplitude et la durée de chaque étape du CSP sont présentées dans le Tableau \ref{tab_4}.

\begin{figure}[!ht]
	\begin{center}
		\includegraphics[width=0.9\columnwidth]{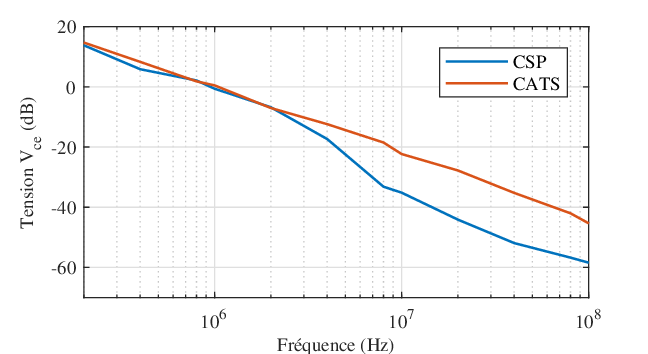}
	\end{center}
		\caption{Résultat de la comparaison à $6,58 W$.}
	\label{fig_10}
\end{figure}

\begin{table}[!ht]
	\caption{Paramètres pour CSP.}
	
	\begin{center}
		\begin{tabular}{|>{\footnotesize}L{1.5cm}|>{\footnotesize}L{1.5cm}|>{\footnotesize}L{1.8cm}|>{\footnotesize}L{1.1cm}|}
			\hline
			\textbf{Paramètre}&\textbf{Valeur}\\
			\hline
	$I_1$& 143,5 mA \\ \hline
	$I_2$ & 14,3 mA \\ \hline
	$I_3$& 20 mA \\ \hline
	$I_4$& 296 mA \\ \hline
	$\Delta t_{1}$& $125,3\mu$s \\ \hline
	$\Delta t_{2}$& $63,3$ ns \\ \hline
	$\Delta t_{3}$& $400$ ns \\ \hline
	$\Delta t_{4}$& $500$ ns \\ \hline
		\end{tabular}
	\end{center}
		\label{tab_4}
\end{table}
Les résultats de la comparaison sont présentés dans \ref{fig_10}. Ils montrent l'enveloppe du spectre de fréquences de CSP et CATS pour le même niveau de pertes par commutations.
Enfin, les résultats de la comparaison montrent que le FOM pour la méthode CATS est 451,97 et pour le CSP est 144,47.

\section{Conclusions}
Dans cet article, une technique de pilotage d'IGBT basée sur un profil de courant de grille est présentée pour réduire les perturbations électromagnétiques. Le CSP proposé nous permet de contrôler le $dv/dt$ par l'amplitude d'un palier ($I_3$) de courant de grille (voire sa durée). La performance du CSP est évaluée par simulation avec différents niveaux de $I_3$, ce qui montre la faisabilité de la méthode de réduction des EMI. Cette technique est comparée à une autre méthode rapportée dans la littérature (CATS) et les résultats de simulation montrent que pour les mêmes pertes par commutations, le CSP génère moins d'EMI. Un inconvénient du procédé proposé, par rapport aux procédés tels que CATS, est qu'il est nécessaire de connaître certains paramètres du transistor pour effectuer un réglage pertinent des autres degrés de liberté de la commande (trois autres paliers de courant à fixer en amplitude et durée). Il est néanmoins envisageable, dans de futures études, de procéder à une adaptation en boucle fermée de ces paramètres sur la base de l'observation des commutations. 

\section{Remerciements}
Les auteurs tiennent à remercier le Conseil national de la science et de la technologie de la science et de la technologie du Mexique (CONACyT) pour la bourse d'études sous le numéro de subvention 705759.


\section{Références}

\begin{thebibliography}{1}

\bibitem{bib_1}{F. Costa et D. Magnon, « Graphical analysis of the spectra of EMI sources in power electronics», in IEEE Transactions on Power Electronics, vol. 20, no. 6, pp. 1491-1498, Nov. 2005.}

\bibitem{bib_2}{A. Galluzzo et al., « Switching characteristic improvement of modern gate controlled devices», 1993 Fifth European Conference on Power Electronics and Applications, Brighton, UK, 1993, pp. 374-379 vol.2.}

\bibitem{bib_3}{A. Consoli et al., «An innovative EMI reduction design technique in power converters», in IEEE Transactions on Electromagnetic Compatibility, vol. 38, no. 4, pp. 567-575, Nov. 1996.}

\bibitem{bib_4}{S. Musumeci et al., «Switching-behavior improvement of insulated gate-controlled devices», in IEEE Transactions on Power Electronics, vol. 12, no. 4, pp. 645-653, July 1997.}

\bibitem{bib_5}{Y. Lobsiger et J. W. Kolar, «Closed-Loop d ii/ d  tt  and d  vv/ d tt  IGBT Gate Driver», in IEEE Transactions on Power Electronics, vol. 30, no. 6, pp. 3402-3417, June 2015.}

\bibitem{bib_6}{H. Riazmontazer et al., «Closed-loop control of switching transition of SiC MOSFETs», 2015 IEEE Applied Power Electronics Conference and Exposition (APEC), Charlotte, NC, USA, 2015, pp. 782-788.}

\bibitem{bib_7}{J. D. Kagerbauer  et T. M. Jahns, «Development of an Active dv/dt Control Algorithm for Reducing Inverter Conducted EMI with Minimal Impact on Switching Losses», 2007 IEEE Power Electronics Specialists Conference, Orlando, FL, USA, 2007, pp. 894-900.}

\bibitem{bib_8}{Shihong Park et T. M. Jahns, «Flexible dv/dt and di/dt control method for insulated gate power switches», Conference Record of the 2001 IEEE Industry Applications Conference. 36th IAS Annual Meeting (Cat. No.01CH37248), Chicago, IL, USA, 2001, pp. 1038-1045 vol.2.}


\bibitem{bib_9}{H. Sawezyn, N. Idir et R. Bausiere, «Lowering the drawbacks of slowing down di/dt and dv/dt of insulated gate transistors», 2002 International Conference on Power Electronics, Machines and Drives (Conf. Publ. No. 487), Sante Fe, NM, USA, 2002, pp. 551-556.}

\bibitem{bib_10}{N. Patin et M. L. Viñals, «Toward an optimal Heisenberg's closed-loop gate drive for Power MOSFETs», IECON 2012 - 38th Annual Conference on IEEE Industrial Electronics Society, Montreal, QC, Canada, 2012, pp. 828-833}

\bibitem{bib_11}{S. Fukunaga, H. Takayama et T. Hikihara, «A Study on Switching Surge Voltage Suppression of SiC MOSFET by Digital Active Gate Drive», 2021 IEEE 12th Energy Conversion Congress  \& Exposition - Asia (ECCE-Asia), Singapore, Singapore, 2021, pp. 1325-1330.}

\bibitem{bib_12}{H. Takayama, T. Okuda et T. Hikihara, «A Study on Suppressing Surge Voltage of SiC MOSFET Using Digital Active Gate Driver», 2020 IEEE Workshop on Wide Bandgap Power Devices and Applications in Asia (WiPDA Asia), Suita, Japan, 2020, pp. 1-5.}

\bibitem{bib_13}{K. Miyazaki et al., «General-purpose clocked gate driver (CGD) IC with programmable 63-level drivability to reduce Ic overshoot and switching loss of various power transistors," 2016 IEEE Applied Power Electronics Conference and Exposition (APEC), Long Beach, CA, USA, 2016, pp. 1640-1645. }

\bibitem{bib_14}{R. Morikawa, T. Sai, K. Hata et M. Takamiya, «New Gate Driving Technique Using Digital Gate Driver IC to Reduce Both EMI in Specific Frequency Band and Switching Loss in IGBTs»,  2020 IEEE 9th International Power Electronics and Motion Control Conference (IPEMC2020-ECCE Asia), Nanjing, China, 2020, pp. 644-651.}

\bibitem{bib_15}{W. Frank, A. Arens et S. Hoerold, «Real-time adjustable gate current control IC solves dv/dt problems in electric drives», PCIM Europe 2014 International Exhibition and Conference for Power Electronics, Intelligent Motion, Renewable Energy and Energy Management, Nuremberg, Germany, 2014, pp. 1-7.}

\bibitem{bib_16}{F. Portuese et M. Melito, «Gate charge leads to easy drive design for Power Mosfet circuits»,  PCIM '90 Proceedings of the 18th International Intelligent Motion Conference, pp. 237-243, 1990.}

\bibitem{bib_17}{Infineon, «Low Loss Duopack: IGBT with Trench and Fieldstop technology IKW40N65ET7», TRENCHSTOP, 2020,Datasheet version 2.2.}

\bibitem{bib_18}{C. R. Paul, «Introduction to Electromagnetic Compatibility», 2nd ed. Hoboken, NJ: Wiley, 2006.}

\bibitem{bib_19}{D. S. Martinez-Padron, N. Patin et E. Monmasson, «Definition and Implementation of an EMI Figure of Merit for Switching Pattern in Power Converters», IECON 2022 – 48th Annual Conference of the IEEE Industrial Electronics Society, Brussels, Belgium, 2022, pp. 1-6.}

\end{thebibliography}
\end{document}